\documentclass[conference]{IEEEtran}
\IEEEoverridecommandlockouts

\usepackage{cite}
\usepackage{times}
\usepackage{latexsym}
\usepackage{algorithm}
\usepackage{algorithmic}
\usepackage{amsmath}
\usepackage{amssymb}
\usepackage{booktabs}
\usepackage{multirow}
\usepackage{array}
\usepackage{caption} %
\usepackage{listings}
\usepackage{xcolor}
\newcommand{\sys}{\textnormal{\textsc{SemaPLC}}}
\usepackage{microtype,inconsolata,xurl}

\usepackage[T1]{fontenc}
\usepackage[utf8]{inputenc}
\usepackage{microtype}
\usepackage{inconsolata}
\usepackage{graphicx}

\definecolor{stkw}{RGB}{0,84,166}
\definecolor{stcomment}{RGB}{124,124,124}
\definecolor{stbg}{RGB}{247,248,250}
\lstdefinelanguage{ST}{
  morekeywords={PROGRAM,END_PROGRAM,FUNCTION_BLOCK,END_FUNCTION_BLOCK,
    VAR,VAR_INPUT,VAR_OUTPUT,END_VAR,AT,
    IF,THEN,ELSE,ELSIF,END_IF,CASE,OF,END_CASE,
    TON,BOOL,INT,DINT,REAL,TIME,TRUE,FALSE,RETAIN},
  sensitive=true, morecomment=[l]{//}, morecomment=[s]{(*}{*)},
  morestring=[b]'
}
\newcommand{\figpass}{%
  \textcolor{green!45!black}{\ensuremath{\checkmark}}}
\newcommand{\figfail}{%
  \textcolor{red!75!black}{\ensuremath{\times}}}

\def\BibTeX{{\rm B\kern-.05em{\sc i\kern-.025em b}\kern-.08em
    T\kern-.1667em\lower.7ex\hbox{E}\kern-.125emX}}
    
\begin{document}
\title{SemaPLC: A Project-Grounded, Verification-Gated Agent Harness for PLC Code Generation\\
\thanks{\IEEEauthorrefmark{1}Corresponding authors.}
}

\author{\IEEEauthorblockN{1\textsuperscript{st} Yanlun Tu}
\IEEEauthorblockA{\textit{Midea AIRC} \\
Shanghai, China \\
tuyl14@midea.com}
\and
\IEEEauthorblockN{2\textsuperscript{nd} Huacan Wang\IEEEauthorrefmark{1}}
\IEEEauthorblockA{\textit{Midea AIRC} \\
Shanghai, China \\
wanghuacan17@mails.ucas.ac.cn}
\and
\IEEEauthorblockN{3\textsuperscript{rd} Jin Xie}
\IEEEauthorblockA{\textit{Midea AIRC} \\
Foshan, China \\
xiejin19@midea.com}
\and
\IEEEauthorblockN{4\textsuperscript{th} Ziyue Zhou}
\IEEEauthorblockA{\textit{Midea AIRC} \\
Shanghai, China \\
ziyue009@e.ntu.edu.sg}
\and
\IEEEauthorblockN{5\textsuperscript{th} Jie Zhou}
\IEEEauthorblockA{\textit{Midea AIRC} \\
Shanghai, China \\
zhoujie195@midea.com}
\and
\IEEEauthorblockN{6\textsuperscript{th} Ningyan Zhu}
\IEEEauthorblockA{\textit{Midea AIRC} \\
Shanghai, China \\
zhuny3@midea.com}
\and
\IEEEauthorblockN{7\textsuperscript{th} Ge Chen}
\IEEEauthorblockA{\textit{KUKA} \\
Foshan, China \\
ge.chen@kuka.com}
\and
\IEEEauthorblockN{8\textsuperscript{th} Wangyi Chen}
\IEEEauthorblockA{\textit{Midea AIRC} \\
Shanghai, China \\
chenwy188@midea.com}
\and
\IEEEauthorblockN{9\textsuperscript{th} Tengfei Zhou}
\IEEEauthorblockA{\textit{KUKA} \\
Shanghai, China \\
frank.zhou@kuka.com}
\and
\IEEEauthorblockN{10\textsuperscript{th} Yifan Zhou}
\IEEEauthorblockA{\textit{SJTU} \\
Shanghai, China \\
yifan-zhou@sjtu.edu.cn}
\and
\IEEEauthorblockN{11\textsuperscript{th} Dasheng Yang\IEEEauthorrefmark{1}}
\IEEEauthorblockA{\textit{ZJU} \\
Hangzhou, China \\
yangdasheng@zju.edu.cn}
\and
\IEEEauthorblockN{12\textsuperscript{th} Xiaofeng Mou}
\IEEEauthorblockA{\textit{Midea AIRC} \\
Shanghai, China \\
mouxf@midea.com}
\and
\IEEEauthorblockN{13\textsuperscript{th} Hui Zhang\IEEEauthorrefmark{1}}
\IEEEauthorblockA{\textit{KUKA} \\
Augsburg, Germany \\
hui.zhang9@kuka.com}
\and
\IEEEauthorblockN{14\textsuperscript{th} Yi Xu\IEEEauthorrefmark{1}}
\IEEEauthorblockA{\textit{Midea AIRC} \\
Shanghai, China \\
xuyi42@midea.com}
}

\maketitle

\begin{abstract}
In industrial practice, AI-generated programmable logic controller (PLC) code must integrate with existing projects and implement the required behavior across scan cycles. Compilation alone does not establish runtime behavior, and subsequent edits can invalidate earlier verification results. We present \sys, a project-grounded, verification-gated agent harness for PLC code generation. The harness combines project context and reusable PLC skills with specification review, compilation, and live-runtime validation. Its completion gate requires recorded checks to meet acceptance criteria for the current implementation. Edits invalidate previous verdicts, and failed checks guide bounded repair. We evaluate \sys\ on 117 independent-POU tasks and 65 project-context tasks across seven language-model backbones. \sys\ outperforms the evaluated baselines on both metrics for every backbone. Its mean verified-pass rate and dynamic score reach 72.6\% and 52.2, exceeding the strongest baseline means by 8.7\% points and 20.8 points, respectively. Verification-layer and failure analyses reveal how build recovery and output observability affect end-to-end results and identify behavioral errors that remain after compilation. We apply the harness in KUKA's iiQworks product environment for PLC development and discuss engineering lessons for its industrial integration and deployment. The industrial use indicates the agent assistance substantially improves PLC development efficiency by 45\%.
\end{abstract}


\begin{figure*}[h!]
\centering
\includegraphics[width=\textwidth]{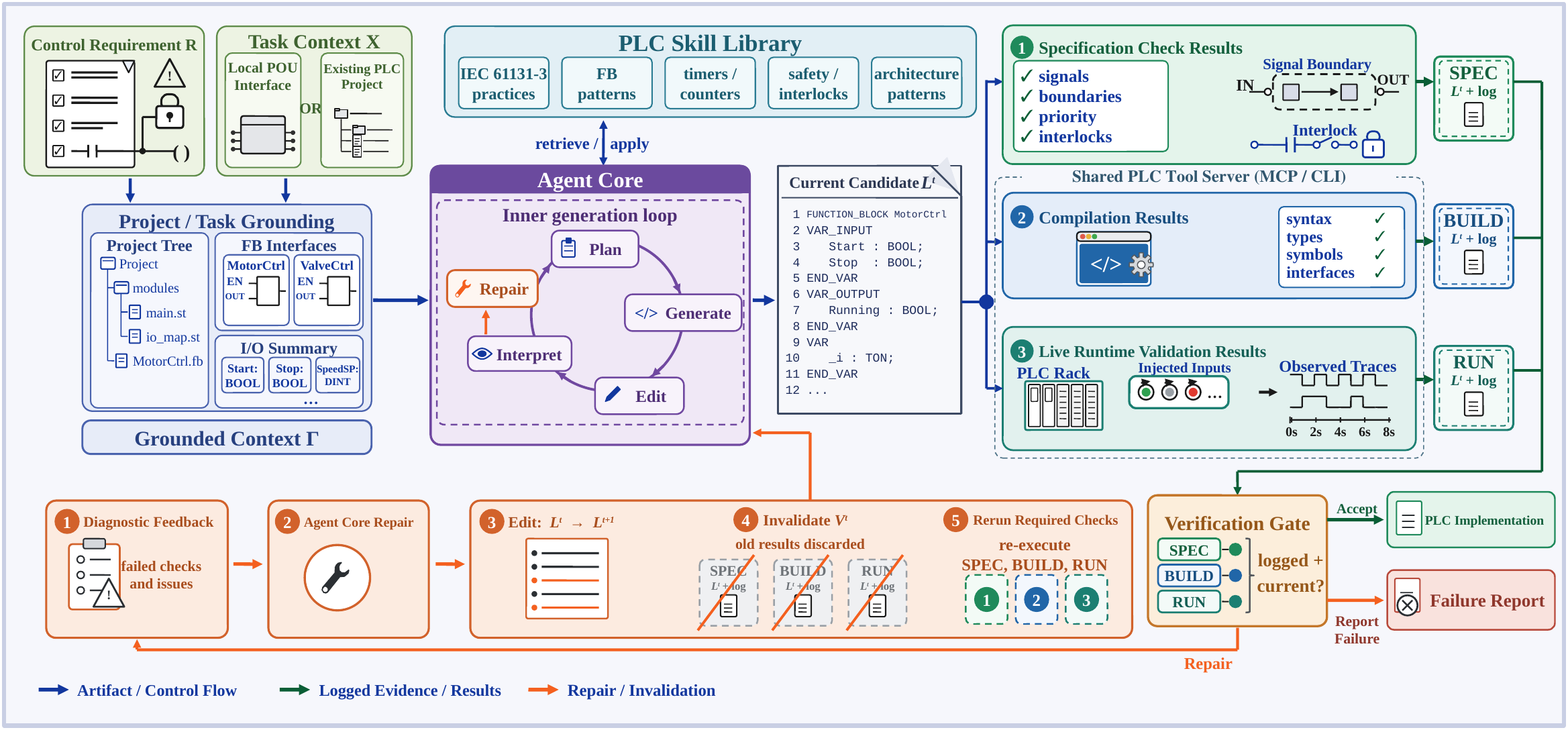}
\caption{Overview of the {\sys}  agent harness:
a model-agnostic agent core uses grounded context, PLC skills,
and a shared PLC tool server (MCP/CLI), while a verification
gate checks current evidence to accept the candidate or
trigger repair.}
\label{fig:arch}
\end{figure*}

\section{Introduction}
\label{sec:intro}
Programmable logic controllers (PLCs) execute industrial control
software~\cite{openplc}. Structured Text (ST), an IEC~61131-3 language,
organizes this logic into program organization units (POUs), including
programs and function blocks~\cite{iec61131}. Control logic must respect
project interfaces, connect signals, and implement the required behavior
across repeated scan cycles. Project integration and behavioral validation are
therefore central to adopting coding agents in industrial development.

Existing PLC-generation systems use compilation, retrieval, formal
properties, or iterative repair~\cite{llm4plc,autoplc,agents4plc}. However, successful compilation and property verification of independent
POUs do not establish whether the generated logic behaves as required
when integrated into an existing project.
Two requirements govern the \emph{AI-assisted PLC development} in industrial
settings like KUKA~\cite{kukaiiqworks}, from requirement interpretation to code validation and delivery. \textbf{(a) Project-grounded
implementation}: generated logic must reuse established declarations and
function-block interfaces, connect the required signals, and follow the
project's execution conventions. \textbf{(b) Verified delivery}: the
implementation must undergo checks of its required behavior, and the
resulting evidence must remain applicable to the delivered source.
An undriven output exposes a project-wiring defect. A wrong setpoint under
competing inputs calls for behavioral validation, while an edit after
validation requires renewed checks. These failure modes motivate
project-aware implementation and verification evidence that remains
current after edits.

\paragraph{Project grounding and evidence-based completion}
We present \sys, a model-agnostic harness that implements these two
requirements in one generation--verification--repair loop. For project
grounding, it retrieves declarations, block interfaces, and entry
conventions before editing the requested logic. For verified delivery,
it combines specification review, integrated compilation, and live-runtime
checks with a binding completion gate. Every reported pass is checked
against its log, an edit invalidates previous verdicts, and a failed check
triggers repair within a bounded budget. The resulting delivery-integrity
property links the delivered implementation to the checks completed for
that version. The agent can explore different repairs while the harness
retains control of the acceptance.

\paragraph{Evaluation and industrial application}
We use two complementary evaluation tracks. The function track assesses
117 independent-POU tasks with a held-out formal judge. The project-context
track assesses 65 tasks adapted from ten industrial plant narratives in
Spec2Control~\cite{spec2control}. Generated logic must compile and execute
inside the supplied ST project. Integrated compilation, static behavior, and dynamic behavior
respectively assess build compatibility, conformance to structural
assertions, and agreement with reference outputs during execution.

Across seven backbones, \sys\ achieves a mean function-level verified-pass
rate of 72.6\% and a mean project-level dynamic score of 52.2 on a 0--100 scale, compared
with the best baseline means of 63.9\% and 31.4. In a single-backbone ablation study, the dynamic behavior score
increases from 23.1 to 54.1 as verification layers are added cumulatively. The accompanying failure breakdown shows fewer
unbuilt executions and missing outputs, while identifying the incorrect
values that remain after execution becomes possible. Together, these
analyses explain how the checks contribute to executable delivery and
where behavioral repair is still needed. %
Cost analysis measures interactions, elapsed time, and inference
expenditure to inform model deployment and serving-cost decisions alongside
the confidentiality requirements of proprietary engineering data.

This paper makes four contributions:
\begin{itemize}
\item \textbf{A project-grounded, verification-gated agent harness.}
\sys\ coordinates specification review, compilation, and live-runtime
validation through a completion rule that binds recorded checks to the
current source. Logged outcomes, edit invalidation, and bounded repair
provide a delivery-integrity mechanism for generated PLC logic.
\item \textbf{Two complementary evaluation tracks.} A 117-task function
track evaluates independent POUs with a held-out formal judge. A 65-task
project-context track evaluates logic integrated into existing ST projects,
reporting integrated compilation, static behavior, and dynamic behavior
separately. Together, they cover local implementation and project-level execution,
using a common scoring protocol for all methods within each track.
\item \textbf{Verification and runtime behavior findings.}
Across three PLC-generation baselines and seven backbones, methods with
similar static behavior scores differ sharply in dynamic behavior scores.
\sys\ achieves the highest verified-pass and dynamic scores on every
backbone. Cumulative ablations, failure diagnostics, and cost measurements
characterize the complete harness's quality--resource trade-offs.
\item \textbf{Industrial integration and deployment experience.}
We integrate the harness into KUKA's iiQWorks product environment and
provide an open-source IDE for research and teaching support. Reports from multiple internal projects indicate an average 45\% reduction in PLC development time with agent assistance. We discuss
self-hosted model selection, engineering-data confidentiality, and how engineers work
with the system for AI-assisted PLC development.
\end{itemize}

\section{Related Work}
\label{sec:related}
\paragraph{LLM-based PLC programming}
LLM4PLC combines ST generation with compiler and model-checker feedback
and user-guided iteration~\cite{llm4plc}. AutoPLC addresses vendor-aware
ST through case retrieval, API recommendation, and compiler-driven
debugging~\cite{autoplc}. Agents4PLC organizes generation and PLCverif
validation in a multiple loop and provides the 117-task benchmark
used in our function track~\cite{agents4plc}. Retrieval-augmented control
code generation also incorporates reusable library
knowledge~\cite{koziolekrag}. Spec2Control evaluates control-logic
engineering from narratives across multiple plants~\cite{spec2control}.
Vendor tools integrate AI assistance with engineering
environments~\cite{siemenscopilot,codesysai}. Our work builds on feedback-driven PLC generation and distinguishes
itself through an explicit artifact--verdict completion rule and a
cross-method evaluation of isolated and project-context tasks.

\paragraph{Agentic code generation and repair}
ReAct interleaves reasoning with tool use~\cite{react}. Self-debugging
and self-refinement exploit execution feedback and generated
critiques~\cite{selfdebug,selfrefine}. SWE-agent studies interfaces for
repository navigation, editing, and testing~\cite{yang2024swe}, InterCode
benchmarks interactive coding with execution feedback~\cite{yang2023intercode},
and CodeT uses model-generated tests to select code~\cite{chen2022codet}.
Sema Code packages agents as reusable infrastructure~\cite{wang2026sema,10.1145/3837729.3840503},
and MCP standardizes tool access~\cite{mcp}. \sys\ applies these
established capabilities to PLC projects, where continuously executing
stateful logic and instrumented runtime tests require explicit tracking
of the checked artifact and its delivery status.

\paragraph{PLC verification and runtime validation}
Model checking has been applied to industrial-sized PLC
software~\cite{modelcheckplc,mcplcsurvey}. PLCverif translates PLC programs
and requirement patterns into model-checker inputs~\cite{plcverif}, and
integration with FRET supports requirement formalization~\cite{fretplcverif}.
Formal verification checks specified properties within the supported
program model. Runtime validation observes behavior under executed inputs.
We use the former in the held-out function judge and the latter for
execution-based repair and project-track scoring. Structural-text
assertions provide the basis for the static behavior score. 

The relationship
to our earlier IDE demonstration~\cite{tu2026semaplc} is discussed in
Section~\ref{sec:reuse}.

\section{Problem Formulation}
\label{sec:problem}
\paragraph{Function track}
Given a natural-language requirement $R_f$ and a local interface $I_f$,
the generator produces a POU $L_f=G(R_f,I_f)$. A held-out judge compiles
the candidate and checks properties derived from $R_f$. Let $V_f$ denote
the fraction of properties verified satisfied. Following the
Agents4PLC~\cite{agents4plc} threshold, the task-level metric is
\begin{equation}
\operatorname{VerifiedPass}(L_f)=\mathbb{1}[V_f\geq0.80].
\label{eq:verifiedpass}
\end{equation}
Unsupported constructs, translation failures, and timeouts produce
inconclusive outcomes, which contribute zero to the satisfied-property
count. Generation failures remain in the task denominator. Verified pass denotes satisfaction
of at least 80\% of the judge's properties under this strict treatment of
inconclusive outcomes. %
We also report compilation success.

\paragraph{Project-context track}
Given requirement $R_p$ and project $P$, the generator produces logic
$L_p=G(R_p,P)$, yielding the integrated program $P'=P\oplus L_p$.
The context includes function-block libraries, declarations, interfaces,
an entry harness, and build configuration. Each task implements a specified
plant section within this project. We report three distinct metrics.
\emph{Integrated compilation} $C(P')\in\{0,1\}$ records whether the
generated logic builds successfully within the supplied project.
\emph{Static behavior} $S(P',R_p)\in[0,100]$ measures the percentage
of satisfied structural assertions on program text, without execution.
\emph{Dynamic behavior} $D(P',\mathcal T)\in[0,100]$ measures agreement
between sampled outputs of the candidate and hidden reference programs
under scenarios $\mathcal T$. All project-level means retain the
complete 65-task denominator.

\paragraph{Generation and completion}
\looseness=-1
For both tracks, $R$ denotes the requirement, $X$ the supplied context,
and $L$ the generated logic. The harness returns $(L,\mathcal V)$ with
recorded check outcomes or reports an incomplete task. Harness completion
uses generation-time checks. Independent evaluation then grades the
candidate with the held-out judge, structural oracle, or golden traces.
Scoring artifacts are isolated from the agent’s generation environment.

\section{The SemaPLC Agent Harness}
\label{sec:method}
\sys\ takes a natural-language control requirement $R$ and task context
$X$, comprising either a local POU interface or an existing PLC project.
The harness configuration supplies the skill library, tool suite, required
checks $K$, per-check repair limit $r$, and interaction budget $B$.
Algorithm~\ref{alg:gate} uses these inputs to generate, verify, and repair
the implementation $L$.

\subsection{Harness Overview}
\sys\ uses a generic event-driven agent core accessed through a
chat-completion interface~\cite{react,wang2026sema}.
Fig.~\ref{fig:arch} organizes the harness into five components:
the agent core, project/task grounding, the PLC skill library,
verification processes, and the verification gate.
The agent plans, generates, edits, and repairs candidates using
project artifacts and skill documents for PLC-specific knowledge.
Verification combines agent-led specification checks using a
developer-authored checklist with compilation and live runtime
validation supported by the shared PLC tool server.
The gate accepts a candidate only when logged evidence for its
current version satisfies all required completion criteria.
Otherwise, diagnostics guide repair within retry and interaction
limits. Candidate changes invalidate prior results and trigger
rechecking. If the criteria remain unmet when these limits are
reached, the harness reports an incomplete task.

To support compilation and runtime validation, the shared PLC tool server exposes three groups of operational capabilities through MCP and CLI interfaces.
\emph{Build and deployment} cover syntax and semantic checking, I/O
discovery, integrated compilation, program loading, and execution control.
\emph{Runtime inspection} provides execution status, diagnostic logs, live
variable access, input forcing, sampled traces, per-scan recordings, and
condition-based waiting. \emph{Behavioral validation} combines input
injection, output assertions, and input release, providing observed
responses for diagnosis and repair.

MCP supports agent-directed interaction, while the CLI supports scripted
verification. Declarative verification plans coordinate compilation,
deployment, input injection, assertions, and cleanup, returning structured
verdicts for the completion gate. This organization gives the agent access
to both individual development operations and repeatable multi-step checks.

\subsection{Project Grounding and PLC Skills}
\textsc{Ground}$(X)$ retrieves the project structure and locates modules
relevant to the requested section. Its output $\Gamma$ comprises the
available variable/type declarations, function-block signatures and
library context, and entry/build conventions needed to implement that
section. Generation reuses these names and interfaces and confines edits
to the requested logic instead of regenerating the project. On the
function track, grounding reduces to the local POU interface.

An unresolved symbol is repaired against the supplied declarations, and
block calls retain established signatures. %
Generation and repair retain project interfaces across iterations.
Compilation checks each candidate against them.

A rules file specifies verification order, tool use, and scan-cycle
conventions. A developer-curated wiki records block signatures, control
patterns, and compiler pitfalls. Procedural skills cover three categories:
requirement review checks candidate logic against the specification,
compiler-guided repair addresses compilation diagnostics, and runtime
validation tests behavior under injected inputs. Each skill defines its
applicability, permitted inputs, prescribed steps, and prohibited actions.
Hidden properties, reference implementations, and scoring traces remain
outside the permitted inputs. Workspace separation enforces this access
boundary (Section~\ref{sec:isolation}).

\subsection{Multi-Source Verification}
\paragraph{Specification review}
The agent applies a developer-authored checklist to the natural-language
requirement and candidate. It checks named devices and signals,
function-block instantiation and per-scan calls, and drivers for published
variables. Boundary checks require explicit strict/non-strict comparisons
and behavior for the remaining branch. Scan-cycle checks distinguish
retained state from values recomputed each scan and examine edge-detection
update order. These checks target semantic omissions that remain
syntactically valid, such as a declared output with no driver.

The review follows each required signal through block invocation to its
published output, checking that values are updated and that competing
branches apply the required output priority within a scan.

The checklist also examines hazardous actuator states during alarms or
interlocks, mutually exclusive commands, regulation on a faulted sensor,
fault/reset fail-safe direction, calibrated engineering ranges, and
permissive bypasses. Each item produces a pass or a concrete revision.
The agent performs this requirement review using the developer-authored
checklist, and the log records its interpretation and result. Compiler
and runtime checks then provide separately computed build outcomes and
execution observations. The independent evaluation uses its own oracles.

\paragraph{Compilation and execution}
Compilation provides externally computed checks of syntax, types,
symbols, interfaces, and the integrated build. Diagnostics identify the
source line and error category. The repair skill addresses the earliest
diagnostic, then recompiles, rather than responding to all cascading
errors with a broad rewrite. Runtime validation supplies a different
kind of evidence: observed outputs under injected inputs. These three
sources share a record format and completion policy, but their evidential
strengths remain distinct.

\subsection{Live Runtime Validation}
The runtime skill deploys an injectable test copy, initializes execution,
forces scenario inputs, and samples external variables. Instrumentation
stays separate from the delivered source. The agent derives scenarios and
expected assertions from the requirement and uses mismatches for repair.
Trusted recordings are also supported, but the experiments withhold scoring
scenarios and golden traces from generation.

Build, deployment, and output checks record different outcomes: accepted
source, a loaded program, and an observed response. The verification plan
combines these stages with cleanup. The atomic behavior check releases
forced inputs after its assertion. Missing observations and recorded
mismatches remain distinct diagnostic outcomes.

The process exercises timer/reset behavior, state evolution, interlocks,
and temporal output selection. Diagnostic guidance associates flat traces
with wiring or undriven outputs, wrong changing values with block logic,
and late transitions with timers or edge detection. These associations
direct the next inspection toward a relevant part of the implementation.
Observable outputs determine the check result, and internal state supports
diagnosis. The benchmark records sampled traces. Persistent assertions
observe retained effects of transient events, while the tool suite also
provides per-scan recording for scan-precise inspection.

Section~\ref{sec:validity} discusses shared requirement-interpretation
errors and sampling coverage. Independent scoring follows the isolation
rules in Section~\ref{sec:isolation}.

\subsection{Verification-Gated Iteration}
Algorithm~\ref{alg:gate} makes completion a harness decision. Each required
check has a recorded verdict and at most $r=2$ repair attempts.
The counters start at zero, increase when their diagnostics are used for
repair, and persist across edits. A changed implementation invalidates
all earlier verdicts. Machine-readable outcome sentinels are cross-checked
against the session's tool-call log. Unsupported success claims become
unchecked outcomes. Missing, failed, or inconclusive required evidence
does not establish completion.

\begin{algorithm}[t]
\caption{Verification-gated generation in SemaPLC}
\label{alg:gate}
\begin{algorithmic}[1]
\REQUIRE requirement $R$, context $X$, required checks $K$ and their
criteria, repair limit $r=2$, budget $B$
\ENSURE $(L,\mathcal V)$ if accepted, otherwise incomplete with $\mathcal V$
\STATE $\Gamma\leftarrow\textsc{Ground}(X)$,
$L\leftarrow\textsc{Generate}(R,\Gamma)$
\STATE $\mathcal V\leftarrow\emptyset$,
$\rho_c\leftarrow0$ for each $c\in K$
\WHILE{budget $B$ remains}
  \FORALL{$c\in K$ without a current verdict}
    \STATE $(v,e_c)\leftarrow\textsc{RunCheck}(c,L)$
    \STATE $\mathcal V[c]\leftarrow v$ if log $e_c$ confirms $v$,
    else \emph{unchecked}
  \ENDFOR
  \IF{all required completion criteria are met}
    \RETURN $(L,\mathcal V)$
  \ENDIF
  \STATE $F\leftarrow\{c\in K\mid c\text{ does not meet its criterion},\;
  \rho_c<r\}$
  \IF{$F=\emptyset$}
    \RETURN incomplete with $\mathcal V$
  \ENDIF
  \STATE $L'\leftarrow\textsc{Repair}(L,\{e_c:c\in F\})$
  \STATE $\rho_c\leftarrow\rho_c+1$ for each $c\in F$
  \IF{$L'\neq L$}
    \STATE $\mathcal V\leftarrow\emptyset$ \COMMENT{invalidate previous checks}
  \ENDIF
  \STATE $L\leftarrow L'$
\ENDWHILE
\RETURN incomplete with $\mathcal V$
\end{algorithmic}
\end{algorithm}

The delivery guard checks that the delivered file contains no located
address literal introduced for testing, that its bytes hash-match the
content included in the last successful compilation, and that the session
records deployment and forced-input evidence. Test instrumentation remains
in the injectable copy, while the guard binds the delivered source to its
successful build and associated test session. Together with edit
invalidation and logged outcomes, this provides \emph{delivery integrity}:
every reported pass applies to the current source and has a supporting
verification record.

Invalidating every verdict avoids relying on the agent to predict the
impact of an edit. Even a small initialization or output-selection change
can affect several checks. Retry counters persist so that modifying the
candidate cannot reset the repair budget or convert incompletion into
acceptance.

\section{Experimental Setup}
\label{sec:setup}
\subsection{Evaluation Tracks and Research Questions}
We use two task settings and four research questions.
\textbf{RQ1 (function-level reliability)} asks how the complete harness
performs against PLC baselines and the same backbone without skills and
tools. \textbf{RQ2 (project-grounded reliability)} asks how generated
logic integrates and executes within an existing project.
\textbf{RQ3 (verification across layers)} examines what compilation,
structural conformance, runtime validation, and the available failure
diagnostics reveal about generated programs. \textbf{RQ4 (interaction
cost)} characterizes requests, elapsed time, and token-priced inference
cost relative to Agents4PLC. The two tracks are complementary evaluations,
not sequential stages of the same task.

\subsection{Task Construction}
\paragraph{Function track}
The 117 independent-POU tasks originate from Agents4PLC~\cite{agents4plc}. PLC engineers conducted an initial three-round audit, identifying oracle defects in 43 tasks. Subsequent review corrected erroneous constants, contradictions, tautologies, unsupported thresholds, and duplicate properties, and clarified underspecified requirements. Overall, 53 tasks were revised, yielding a benchmark with 607 verification properties. All methods are evaluated on the same repaired benchmark and held-out judge.

\paragraph{Project-context track}
We adapt 65 tasks from the ten industrial plant narratives in
Spec2Control~\cite{spec2control} into ST projects. Each task supplies a
section narrative, function-block interfaces and library context, and an
empty entry harness. The generated logic must build and execute within
the full project. Reference implementations, original answers, and
scoring traces are withheld from every method. 
No verbatim reference copies were found in task packages or run directories.

\subsection{Models and Baselines}
Both tracks use seven backbones from five vendors:
MiniMax-M2.7~\cite{minimaxm2}, MiniMax-M3~\cite{minimaxm3},
Qwen3.5-Plus~\cite{qwen35}, DeepSeek-V4-Flash and
DeepSeek-V4-Pro~\cite{deepseekv4}, GLM-5.2~\cite{glm5}, and
GPT-5.5~\cite{gpt55}. For each backbone, all methods call the same
endpoint. We compare LLM4PLC~\cite{llm4plc}, AutoPLC~\cite{autoplc}, and Agents4PLC~\cite{agents4plc}. 
Baselines retain their reported configurations. The \emph{bare}
configuration removes skills and tools from the same SemaPLC agent loop,
so its comparison with \emph{full} measures the combined harness.

All candidates use common build and scoring infrastructure, while methods
retain their feedback and iteration policies. This evaluates complete
configurations on one ST environment with the same task inputs and output
criteria. Section~\ref{sec:validity} discusses iteration budgets, toolchain
adaptation, and run variability.

\subsection{Metrics and Verification Infrastructure}
\paragraph{Function track}
RuSTy compiles each candidate~\cite{rusty}. PLCverif with the nuXmv
backend checks requirement-derived properties~\cite{plcverif}.
Equation~\ref{eq:verifiedpass} determines verified pass.
Inconclusive properties and empty generations remain failures in the
reported rate. The coverage diagnostic in Section~\ref{sec:layers}
separates verifier conclusiveness from satisfaction.

\paragraph{Integrated compilation and static behavior}
The project metric $C$ is a binary RuSTy build of $P'$. The static score
$S$ is the percentage of satisfied structural assertions evaluated
without execution: substring presence/absence, required calls, normalized
numeric constants, structural references, and any-of sets. For 63 tasks,
a parser extracts string and numeric literals from section-specific corpus
runtime tests. An assertion is retained only if the hidden reference
satisfies it. Every oracle must give its reference full marks. The
remaining two tasks use manually written oracles validated in the same
way. There are 502 assertions in total, with 1--20 per task and a median
of seven. Validation against the reference ensures consistency between
an oracle and its intended implementation. This metric records structural
conformance to the selected assertions.

\paragraph{Dynamic behavior}
Candidate and reference execute in identical runtime harnesses under up
to six scenarios per task. The scenario set contains normal operation
from the corpus runtime project, limit violations for analog inputs with
configured high/low bounds, and a discrete-flip scenario that inverts
Boolean inputs. A limit violation offsets the engineering value beyond
the configured bound and inverse-scales it to the raw input signal.
The observed ports are the reference function block's outputs. Ports
matching alarm/fault naming patterns are reported but excluded from the
score. The metric still scores process outputs under fault-related inputs.

Each scenario scores the fraction of core output-port traces matching
the reference, using exact equality for Boolean and numeric samples.
The task score $D$ is the mean scenario score, and reported project means
use all 65 tasks. Compilation, deployment, timeout, or missing-trace
failure on either side receives zero. The metric therefore measures
\emph{end-to-end trace agreement}, combining executable delivery with
matching output traces across the complete task set.

Structural assertions can score noncompiling source, whereas $D$ requires
execution. Repairing a build can therefore improve $D$ by making behavior
observable. Reporting $C$, $S$, and $D$ separately distinguishes project
compatibility, source conformance, and execution outcomes. Exact sample
matching uses the shared runtime configuration defined above.
Section~\ref{sec:validity} discusses its coverage and transfer across
execution environments.

\subsection{Evaluation Isolation and Information Access}
\label{sec:isolation}
A generation workspace contains only task inputs, interface/library
context, and the target file. Held-out function properties, structural
oracles, reference implementations, and golden traces are absent.
Every method has the same access restriction. Compiler diagnostics are
available during generation and also determine integrated compilation.
The agent's own audit and runtime feedback guide repair but contribute
no points to the final semantic scores.

Generation-time scenarios and expected assertions come from the
natural-language requirement. Evaluation scenarios are constructed
separately from the corpus and hidden reference. Both address the same
control requirement, while workspace separation keeps scoring artifacts
inaccessible throughout generation. Section~\ref{sec:reuse} describes
software reuse and the public implementation.

\begin{table*}[t]
\caption{Function-track strict verified-pass rate (\%, 117 tasks).
The shared held-out judge treats inconclusive properties as unsatisfied
and empty generations as failures. Bare removes skills and tools.
Full uses the complete harness.
DS = DeepSeek.}
\label{tab:main}
\centering
\small
\setlength{\tabcolsep}{8pt}
\begin{tabular}{l rrrrr}
\toprule
& & & & \multicolumn{2}{c}{\textsc{SemaPLC}} \\
\cmidrule(lr){5-6}
Model & LLM4PLC & AutoPLC & Agents4PLC & bare & full \\
\midrule
MiniMax-M2.7 & 22.2 & 49.6 & 53.8 & 39.3 & \textbf{69.2} \\
MiniMax-M3   & 15.4 & 65.0 & 55.6 & 60.7 & \textbf{69.2} \\
Qwen3.5-Plus & 13.7 & 67.5 & 67.5 & 62.4 & \textbf{75.2} \\
DS-V4-Flash  & 41.0 & 54.7 & 54.7 & 34.2 & \textbf{67.5} \\
DS-V4-Pro    & 43.6 & 61.5 & 62.4 & 55.6 & \textbf{69.2} \\
GLM-5.2      & 30.8 & 59.0 & 74.4 & 63.2 & \textbf{76.1} \\
GPT-5.5      & 44.4 & 79.5 & 78.6 & 71.8 & \textbf{82.1} \\
\midrule
Mean         & 30.2 & 62.4 & 63.9 & 55.3 & \textbf{72.6} \\
Worst        & 13.7 & 49.6 & 53.8 & 34.2 & \textbf{67.5} \\
\bottomrule
\end{tabular}

\end{table*}

\section{Results}
\label{sec:results}
\subsection{RQ1: Function-Level Reliability}
Table~\ref{tab:main} applies the same held-out verified-pass criterion to
all methods. \sys\ obtains the highest rate on every backbone. Its mean
of 72.6\% exceeds Agents4PLC's 63.9\% by 8.7 percentage points. On GPT-5.5,
\sys\ reaches 82.1\%, compared with the best baseline rate of 79.5\%.
Its minimum across backbones is 67.5\%, and its cross-model range is
14.6 points. These results show that the complete harness improves
function-level verification across the evaluated model families.

\paragraph{Contribution of the complete harness}
Every backbone improves over its bare configuration, by 8.5--33.3 points.
The largest gains occur for MiniMax-M2.7 ($+29.9$) and
DeepSeek-V4-Flash ($+33.3$). The mean gain is 17.3 points, and the
cross-model spread contracts from 37.6 to 14.6 points. Mean compilation
also increases from 85.5\% to 99.2\%. The comparison measures the combined
benefit of domain skills, tool access, and verification-guided iteration.
The largest improvement and highest final score occur on different
backbones: DeepSeek-V4-Flash gains 33.3 points, while GPT-5.5 finishes at
82.1\%. Model--harness selection therefore considers both improvement over
bare and final quality. 

\subsection{RQ2: Project-Grounded Reliability}
Table~\ref{tab:project} separates the three project-level outcomes.
\sys\ obtains a mean integrated-compilation rate of 89.4\%, compared
with 58.7--81.5\% for the baselines. It also obtains the highest mean
static behavior score, 81.6, and leads this metric on five of seven backbones.
The dynamic mean is 52.2, compared with 31.4 for AutoPLC, 30.3 for
Agents4PLC, and 22.4 for LLM4PLC. Thus the gain over the strongest
project-dynamic baseline is 20.8 points. \sys\ leads dynamic scoring
on every backbone, while its lowest score is 31.3 rather than the
single-digit minima observed for all three baselines.

The three metrics reveal where a project candidate succeeds in the
development process. Integrated compilation measures compatibility with
the supplied project, while the dynamic behavior score measures
agreement with reference outputs during execution. The dynamic mean of 52.2 includes compilation and
execution failures as well as mismatched values. 

The improvement varies across backbones. On GPT-5.5, \sys\ reaches 65.4
dynamically against Agents4PLC's 63.6, while its static score is 84.1
against baseline values of 86.3--88.8. %
The static and dynamic behavior scores rank the same model--harness
configurations differently, distinguishing structural conformance from
runtime output agreement.

A second contrast appears with DeepSeek-V4-Pro. AutoPLC compiles more
project candidates than SemaPLC (95.4\% versus 89.2\%), yet its dynamic
score is lower (45.7 versus 57.4). Thus, the observed dynamic ordering is
not simply the ordering of build success. The same task set and backbone
favor different methods under compilation and executed-output criteria. This comparison reinforces the need
to keep integration and behavior as separate acceptance dimensions:
a higher compile rate alone can therefore favor a development
configuration with a lower dynamic behavior score.

\begin{table*}[t]
\caption{Project-context results (0--100, all 65 tasks): integrated
compilation, static behavior, and dynamic behavior.
Bold marks the highest value per column. Worst/Best/Mean summarize the
seven backbones.}
\label{tab:project}
\centering
\small
\setlength{\tabcolsep}{3pt}
\begin{tabular*}{\textwidth}{@{\extracolsep{\fill}}l rrrrrrr rrr@{}}
\toprule
& MiniMax & MiniMax & Qwen3.5 & DeepSeek & DeepSeek & & & & & \\
Method & M2.7 & M3 & Plus & V4-Flash & V4-Pro & GLM-5.2 & GPT-5.5 & Worst & Best & Mean \\
\midrule
\multicolumn{11}{l}{\itshape Integrated compilation} \\
LLM4PLC            & 47.7 & 53.8 & 16.9 & 60.0 & 52.3 & 80.0 & \textbf{100.0} & 16.9 & \textbf{100.0} & 58.7 \\
AutoPLC            & 69.2 & \textbf{95.4} & 58.5 & 58.5 & \textbf{95.4} & \textbf{95.4} & 98.5 & 58.5 & 98.5 & 81.5 \\
Agents4PLC         & 47.7 & 69.2 & 40.0 & 75.4 & 78.5 & 89.2 & 98.5 & 40.0 & 98.5 & 71.2 \\
\textsc{SemaPLC}   & \textbf{81.5} & \textbf{95.4} & \textbf{80.0} & \textbf{84.6} & 89.2 & \textbf{95.4} & \textbf{100.0} & \textbf{80.0} & \textbf{100.0} & \textbf{89.4} \\
\addlinespace
\multicolumn{11}{l}{\itshape Static behavior} \\
LLM4PLC            & \textbf{76.1} & 76.7 & 74.5 & 70.3 & 69.2 & 77.1 & 86.3 & 69.2 & 86.3 & 75.7 \\
AutoPLC            & 68.9 & 73.5 & 68.9 & 76.2 & 66.5 & 75.5 & \textbf{88.8} & 66.5 & \textbf{88.8} & 74.0 \\
Agents4PLC         & 63.4 & 73.5 & 77.6 & 71.4 & 64.5 & 62.8 & 88.6 & 62.8 & 88.6 & 71.7 \\
\textsc{SemaPLC}   & 74.9 & \textbf{84.9} & \textbf{79.9} & \textbf{78.0} & \textbf{81.2} & \textbf{88.0} & 84.1 & \textbf{74.9} & 88.0 & \textbf{81.6} \\
\addlinespace
\multicolumn{11}{l}{\itshape Dynamic behavior} \\
LLM4PLC            & 3.0 & 26.1 & 6.6 & 18.4 & 13.9 & 34.6 & 54.5 & 3.0 & 54.5 & 22.4 \\
AutoPLC            & 4.0 & 43.5 & 19.8 & 23.9 & 45.7 & 21.9 & 61.1 & 4.0 & 61.1 & 31.4 \\
Agents4PLC         & 4.5 & 30.8 & 11.4 & 28.3 & 28.8 & 44.6 & 63.6 & 4.5 & 63.6 & 30.3 \\
\textsc{SemaPLC}   & \textbf{31.3} & \textbf{52.1} & \textbf{43.1} & \textbf{54.1} & \textbf{57.4} & \textbf{61.9} & \textbf{65.4} & \textbf{31.3} & \textbf{65.4} & \textbf{52.2} \\
\bottomrule
\end{tabular*}
\end{table*}

\subsection{RQ3: Verification across Layers}
\label{sec:layers}
\paragraph{Different checks expose different failures}
The baseline static means occupy a four-point band (71.7--75.7), but
their dynamic means span 22.4--31.4 and have a different ranking:
LLM4PLC has the highest baseline static score and the weakest
dynamic score. For GPT-5.5, \sys\ has the lowest static score
and the highest dynamic score. These reversals show that this
structural-text oracle and runtime execution distinguish different
properties of the generated programs. The practical implication is to
retain both source-level conformance and output-level execution checks
when evaluating project-level PLC generation.

\paragraph{Case study}
Fig.~\ref{fig:case-runtime} traces the coking-refinery example from plant section~8.
The requirement assigns a low-flow setpoint of 500 and a transmitter-fault
setpoint of 2500, with fault taking priority. LLM4PLC and AutoPLC fail
compilation. Agents4PLC compiles, but its second assignment overwrites the
fault default when low flow also holds, as shown by source analysis.
\sys's intermediate candidate also compiles, yet passes one fixed manual
setpoint to the control block for both causes. Forcing low flow produces
2500 instead of the required 500. The runtime mismatch localizes the
repair to cause-specific manual-output selection. The agent preserves the
existing block interface, revises the selected argument, and reruns
compilation and runtime checks. Re-verification confirms the displayed
low-flow and fault cases. The case connects project grounding, executable
feedback, and edit invalidation in one complete repair sequence.

\begin{figure*}[t]
\centering
\begin{minipage}{\textwidth}
\centering

\fcolorbox{black}{gray!8}{%
\parbox{\dimexpr\linewidth-2\fboxsep-2\fboxrule\relax}{\centering\footnotesize
\textbf{Case requirement:}\quad
Low flow without fault ($FT$-701 $<50$\,kg/hr)
$\Rightarrow$ SlaveSP $=500$
\qquad
Transmitter fault (overrides low flow)
$\Rightarrow$ SlaveSP $=2500$}}
\par\vspace{3pt}

\begin{minipage}[t]{0.485\linewidth}
\centering

{\setlength{\fboxsep}{3.5pt}%
\fcolorbox{red!65!black}{red!3}{%
\parbox{\dimexpr\linewidth-2\fboxsep-2\fboxrule\relax}{\centering\footnotesize\bfseries
Baselines: compilation or output-priority defects}}}
\par\vspace{3pt}

{\scriptsize\bfseries
LLM4PLC \figfail\ Compile failure
\quad
AutoPLC \figfail\ Compile failure}

\begin{lstlisting}[
  aboveskip=2pt,
  belowskip=0pt,
  xleftmargin=\dimexpr2pt+\fboxrule\relax,
  xrightmargin=\dimexpr2pt+\fboxrule\relax,
  basicstyle=\ttfamily\fontsize{7}{7.6}\selectfont
]
(* Rejected by the RuSTy syntax/semantic gate;
   neither candidate reaches the runtime. *)
LLM4PLC  E007  expected ';' before RATIO_SP (VAR block)
AutoPLC  E048  unresolved ref PT_701_High_Alarm
\end{lstlisting}

\vspace{4pt}

{\scriptsize\bfseries
Agents4PLC \figpass\ Compiles
\quad
\figfail\ Priority defect (source analysis)}

\begin{lstlisting}[
  aboveskip=2pt,
  belowskip=0pt,
  xleftmargin=\dimexpr2pt+\fboxrule\relax,
  xrightmargin=\dimexpr2pt+\fboxrule\relax,
  basicstyle=\ttfamily\fontsize{7}{7.6}\selectfont
]
(* Fault first: assign the required default. *)
IF Fault_Active THEN
    Air_Flow_SP := 2500.0;
END_IF;

(* Separate IF also runs under fault *)
IF FT_701_PV < 50.0 THEN
    Air_Flow_SP := 500.0;  (* overwrites 2500 *)
END_IF;
\end{lstlisting}
\end{minipage}\hfill%
\begin{minipage}[t]{0.485\linewidth}
\centering

{\setlength{\fboxsep}{3.5pt}%
\fcolorbox{green!45!black}{green!3}{%
\parbox{\dimexpr\linewidth-2\fboxsep-2\fboxrule\relax}{\centering\footnotesize\bfseries
SemaPLC:
detect $\rightarrow$ repair $\rightarrow$ verify}}}
\par\vspace{3pt}

{\scriptsize\bfseries
1. Intermediate candidate \figpass\ Compiles}

\begin{lstlisting}[
  aboveskip=2pt,
  belowskip=0pt,
  xleftmargin=\dimexpr2pt+\fboxrule\relax,
  xrightmargin=\dimexpr2pt+\fboxrule\relax,
  basicstyle=\ttfamily\fontsize{7}{7.6}\selectfont
]
(* One static manual output serves both cases. *)
FFIC_701(Auto := Permissives AND
  (FT_701.PV >= 50.0), MV_MIN := 0.0, MV_MAX := 5000.0,
  ManOut := 2500.0, SlaveSP => FFIC701_SlaveSP);
\end{lstlisting}

\vspace{4pt}

\fcolorbox{orange!80!black}{orange!4}{%
\parbox{\dimexpr\linewidth-2\fboxsep-2\fboxrule\relax}{%
\scriptsize\ttfamily\raggedright
\textcolor{orange!80!black}{%
\textbf{2. Runtime Verification \figfail}\quad
Force $FT$-701 $=30$ kg/hr.\\[-0.2ex]
Expected SlaveSP $=500$, observed $2500$.\\[-0.2ex]
The low-flow minimum is never applied.}}}

\vspace{4pt}

{\scriptsize\bfseries
3. Repair and re-verify \figpass\ Both cases pass}

\begin{lstlisting}[
  aboveskip=2pt,
  belowskip=0pt,
  xleftmargin=\dimexpr2pt+\fboxrule\relax,
  xrightmargin=\dimexpr2pt+\fboxrule\relax,
  basicstyle=\ttfamily\fontsize{7}{7.6}\selectfont
]
(* Select the manual output by cause. *)
IF Fault_Active THEN FFIC701_ManOut := 2500.0;
ELSE FFIC701_ManOut := 500.0;
END_IF;
FFIC_701(..., ManOut := FFIC701_ManOut, ...);
\end{lstlisting}
\end{minipage}

\vspace{4pt}

\begin{minipage}[t]{0.485\linewidth}
\centering
\fcolorbox{red!65!black}{red!3}{%
\parbox[c][13pt][c]{\dimexpr\linewidth-2\fboxsep-2\fboxrule\relax}{\centering\footnotesize\normalfont
\figfail\ Fault + low flow: 500 (source-inferred)
}}
\end{minipage}\hfill%
\begin{minipage}[t]{0.485\linewidth}
\centering
\fcolorbox{green!45!black}{green!3}{%
\parbox[c][13pt][c]{\dimexpr\linewidth-2\fboxsep-2\fboxrule\relax}{\centering\footnotesize\normalfont
\figpass\ Low flow = 500
\hspace{0.35em}
\figpass\ Fault = 2500
\hspace{0.35em}
\figpass\ Runtime verified}}
\end{minipage}

\vspace{4pt}

\fcolorbox{green!45!black}{green!5}{%
\parbox{\dimexpr\linewidth-2\fboxsep-2\fboxrule\relax}{\centering\footnotesize\bfseries
Runtime evidence exposes a compiling defect and directs a cause-specific repair.}}
\end{minipage}

\caption{Project-track case: section~8 of the coking-refinery plant.
Transmitter fault takes priority over low flow.
LLM4PLC and AutoPLC fail compilation. Agents4PLC compiles, but sequential
assignments overwrite the fault default when low flow also holds. Its 500
output is inferred from source, not measured in a runtime trace.
For SemaPLC, the observed low-flow mismatch (500 expected, 2500 observed)
triggers repair, and re-verification confirms the two displayed cases.
Identifiers are method-specific. Ellipses
omit unchanged arguments.}
\label{fig:case-runtime}
\end{figure*}

\paragraph{Cumulative verification-layer ablation}
Table~\ref{tab:projablation} adds specification review, compilation,
and runtime validation in that order on DeepSeek-V4-Flash. Dynamic
scores increase from 23.1 to 30.3, 43.7, and 54.1, while the static
score increases from 71.5 to 78.0. The compilation step gives a
13.4-point dynamic increase. Build failures score zero at runtime,
so this increment includes recovering executable candidates. Adding
runtime validation then raises compilation from 83.1\% to 84.6\%, but
raises the dynamic score by 10.4 points. This separation shows the
additional value of evaluating outputs after most candidates already
build.

Interaction increases from 34k tokens and 8.9 requests per task to 129k
and 47.8. Runtime validation has the largest interaction increment in
this ordering, reflecting its build, deployment, input-injection, and
observation steps. Together with the 10.4-point dynamic improvement over
specification review plus compilation, this identifies a concrete
quality--resource trade-off for adding execution feedback. 

\begin{table}[t]
\caption{Cumulative verification-layer ablation on DeepSeek-V4-Flash
(65 project tasks). Comp./Stat./Dyn. denote integrated compilation,
static behavior, and dynamic behavior, respectively, as in
Table~\ref{tab:project}.
Tok./Reqs. are mean tokens and model requests per task.}
\label{tab:projablation}
\centering
\small
\setlength{\tabcolsep}{3.5pt}
\begin{tabular}{l ccc cr}
\toprule
Verification layers & Comp. & Stat. & Dyn. & Tok. & Reqs. \\
\midrule
None (generate only) & 64.6 & 71.5 & 23.1 & 34k & 8.9 \\
$+$\,Spec            & 70.8 & 74.0 & 30.3 & 60k & 14.6 \\
$+$\,Compile         & 83.1 & 77.8 & 43.7 & 74k & 25.5 \\
$+$\,Runtime (full)  & \textbf{84.6} & \textbf{78.0} & \textbf{54.1} & 129k & 47.8 \\
\bottomrule
\end{tabular}
\end{table}

\paragraph{Execution availability and residual wrong values}
Table~\ref{tab:checkoutcome} summarizes outcomes for the cumulative
configurations. The combined share of not-built/run and missing-port
outcomes falls from 74.7\% to 21.5\%. In the full configuration, these
categories account for 3.3\% and 18.2\%, respectively. This distinction
separates build and execution failures from missing outputs in running
programs.

Recorded wrong values rise from 2.1\% to 24.4\%, while correct outcomes
increase from 23.1\% to 54.1\%. Cases that do not build or run provide no runtime observations for assessing output correctness. As execution availability increases, a larger recorded wrong-value share alone does not indicate worse generated logic. Normal-operation and
limit-breach mismatches account for 7.1\% and 17.3\%, respectively.

Missing ports direct inspection toward interface mapping and output
wiring, while wrong values direct it toward logic under the recorded
scenario. Section~\ref{sec:lessons} develops these implications for
engineering review.

\begin{table}[t]
\caption{Outcome breakdown for the cumulative configurations
(\%, DeepSeek-V4-Flash, 65 project tasks). Normal/breach subrows contribute to the overall
distribution. Subrow sums differ from totals because of rounding.}
\label{tab:checkoutcome}
\centering
\begin{tabular}{l rrrr}
\toprule
Outcome & None & $+$Spec & $+$Compile & $+$Runtime \\
\midrule
Check failed              & 76.9 & 69.7 & 56.3 & 45.9 \\
\quad not built/run       & 33.4 & 24.5 & 19.7 & 3.3  \\
\quad port missing        & 41.3 & 35.7 & 20.0 & 18.2 \\
\quad wrong value (normal) & 0.9  & 3.0  & 4.9  & 7.1  \\
\quad wrong value (breach) & 1.2  & 6.5  & 11.7 & 17.3 \\
\midrule
Correct                   & 23.1 & 30.3 & 43.7 & 54.1 \\
\bottomrule
\end{tabular}

\end{table}

\paragraph{Coverage of the held-out formal pipeline}

Table~\ref{tab:coverage} reports the coverage of the evaluated
PLCverif/nuXmv pipeline in a separate diagnostic pool of 1{,}293 delivered
SemaPLC programs. Conclusive outcomes establish either satisfaction or
violation of a property. They account for 75.7\% of properties in the
group without \texttt{REAL}/timers and 87.0\% in the \texttt{REAL} group.
All 174 properties for the 32 \texttt{TON}-bearing programs are
inconclusive, with unsupported patterns and translation failures
dominating. These results expose a coverage limitation for timer-bearing
programs in this pipeline and diagnostic pool.

An inconclusive outcome does not establish a behavioral violation.
However, Equation~\ref{eq:verifiedpass} counts it as a failure, so the
strict verified-pass rate reflects both generated behavior and verifier
support. The rate therefore measures success under the evaluated formal
pipeline, whose coverage limits must inform interpretation of failed
tasks. Runtime validation can provide complementary execution evidence
for timer and stateful behavior within the exercised scenarios.

\begin{table}[t]
\caption{Formal-verification coverage pooled over 1{,}293 delivered
SemaPLC programs. This property-coverage diagnostic is separate from the
117-task accuracy calculation. Conclusive means satisfied or violated.}
\label{tab:coverage}
\centering
\small
\setlength{\tabcolsep}{4pt}
\begin{tabular}{l rrrr}
\toprule
Program group & Programs & Props. & Concl.\,\% & Inconcl.\,\% \\
\midrule
No \texttt{REAL}/timer     & 916 & 4653 & 75.7 & 24.3 \\
With \texttt{REAL}         & 345 & 1845 & 87.0 & 13.0 \\
With timer (\texttt{TON})  & 32  & 174  & 0.0  & 100.0 \\
\bottomrule
\end{tabular}

\end{table}

\subsection{RQ4: Interaction Cost}
Table~\ref{tab:cost} compares SemaPLC with Agents4PLC. Request values
average per-task means across backbones, and time values average the
corresponding per-backbone medians. Parentheses report cross-backbone
ranges. Token-priced inference cost applies the relevant rates to
recorded uncached-input, cached-input, and output usage, normalized to
Agents4PLC within each track.

For backbone $m$, the accounting can be written as
\begin{equation}
\operatorname{Cost}_m=\sum_j
\left(p_m^{u}u_j+p_m^{h}h_j+p_m^{o}o_j\right),
\label{eq:tokencost}
\end{equation}
where $j$ indexes recorded requests, $u_j$, $h_j$, and $o_j$ denote
uncached-input, cached-input, and output tokens, and the corresponding
$p_m$ values are their per-token rates. Cached tokens are counted separately from uncached input. Expenditure
therefore depends on token categories and prices, not request count alone.

On the function track, requests are comparable (6.5 versus 6.3), while
the reported elapsed-time aggregate is 71 versus 454 seconds and
SemaPLC's inference cost is $0.52\times$. Agents4PLC invokes PLCverif
and nuXmv inside its iterative workflow, contributing substantial
verification latency. On the project track, elapsed-time aggregates
are similar (347 versus 344 seconds), but SemaPLC uses 34.1 requests
against 6.9. The tool-driven loop requires more interactions for
deployment, input manipulation, observation, and repair.

On the project track, SemaPLC uses approximately $4.9\times$ as many
requests as Agents4PLC, with an inference-cost ratio of $1.49\times$.
Request lengths and input-cache utilization explain why interaction
count and monetary cost differ. Retained conversation history increases
both context size and opportunities for reuse. The observed configuration
therefore pairs a 52.2 dynamic score, against 30.3 for Agents4PLC, with
49\% higher token-priced expenditure.

Project-track per-backbone median times range from 25 to 1380 seconds
for SemaPLC and 47 to 917 for Agents4PLC. Mean request counts range from
16.4 to 60.4 and 6.8 to 7.0, respectively. %
These ranges show variation in repair-loop latency and interaction across
backbones. Similar cross-backbone time aggregates should therefore be read
alongside each backbone's achieved quality and resource use.

\begin{table}[t]
\caption{Interaction cost relative to Agents4PLC. Values average
per-backbone request means and time medians. Parentheses give
cross-backbone ranges. Relative token-priced cost includes uncached
input, cached input, and output usage, normalized within each track.}
\label{tab:cost}
\centering
\small
\setlength{\tabcolsep}{3pt}
\renewcommand{\arraystretch}{1.15}
\begin{tabular*}{\linewidth}{@{\extracolsep{\fill}}l ccc@{}}
\toprule
Method & Requests / task & Time / task (s) & Token cost \\
\midrule
\multicolumn{4}{l}{\itshape Function track} \\
Agents4PLC       & 6.3 \ (4.4 -- 7.2) & 454 \ (241 -- 688) & $1\times$ \\
\textsc{SemaPLC} & 6.5 \ (5.5 -- 7.6) & 71 \ (41 -- 156) & $0.52\times$ \\
\addlinespace
\multicolumn{4}{l}{\itshape Project track} \\
Agents4PLC       & 6.9 \ (6.8 -- 7.0) & 344 \ (47 -- 917) & $1\times$ \\
\textsc{SemaPLC} & 34.1 \ (16.4 -- 60.4) & 347 \ (25 -- 1380) & $1.49\times$ \\
\bottomrule
\end{tabular*}

\end{table}

\section{Discussion}
\label{sec:discussion}
The study identifies design choices for taking AI-assisted PLC programming
from generated source to a checked project implementation. We relate the
experimental findings to verification-aware engineering, internal model
deployment, and reuse through open-source and product-specific editions.

\subsection{Engineering Lessons for Agent Harnesses}
\label{sec:lessons}
\paragraph{Bind completion to the current artifact}
A verification result has a source version and a check context. Tracking
both makes the implementation and its current verification record a
single engineering handoff. After an edit, the gate requires renewed
checks before accepting the candidate. An incomplete outcome identifies
the unresolved checks and directs the next repair or human review.
The design separates the agent's freedom to choose edits from the
harness's responsibility to enforce acceptance criteria.

\paragraph{Treat observability as part of integration}
Table~\ref{tab:checkoutcome} leaves 18.2\% of checks without a required
port even though only 3.3\% lack a built/running candidate. Project
integration therefore includes exposing the outputs that the test harness
must observe. Missing ports identify interface-mapping or wiring work,
while recorded wrong values identify behavior to inspect under a concrete
scenario. This distinction helps engineers allocate repair effort,
particularly once most build failures are resolved.

\paragraph{Preserve the role of each check}
Specification review maps requirements to candidate logic, compilation
checks project compatibility, and runtime validation observes executed
outputs. A shared completion policy coordinates these distinct checks.
In Fig.~\ref{fig:case-runtime}, the agent retains the control-block
interface while changing the argument selected under low flow. The
successful repair relies on both interface preservation and an observed
output mismatch. Repeating the checks after the edit connects this
localized repair to the final delivery decision.

\subsection{Deployment and Resource Trade-offs}
\label{sec:deployment}
Industrial deployment balances generated-code quality, operating cost,
and confidentiality of engineering data. In our internal deployment,
we use self-hosted DeepSeek-V4-Flash and DeepSeek-V4-Pro for
AI-assisted PLC development. Their openly available weights support
local deployment~\cite{deepseekv4flashweights,deepseekv4proweights}.
This choice combines control over inference infrastructure with the
ability to process proprietary project context inside the company.
Source code, variable declarations, and control requirements can then
be supplied to the model without sending the inference request to an
external model service.

The project-track results provide a quality reference: with SemaPLC,
DeepSeek-V4-Flash and DeepSeek-V4-Pro achieve dynamic scores of 54.1 and
57.4. Internal model selection combines this effectiveness with serving
cost and self-hosting availability. RQ4's token-priced accounting measures
API expenditure. Self-hosted operation additionally requires accounting
for hardware capacity, utilization, and maintenance.

Internal endpoints use the same grounding and verification procedure.
Configurable resource budgets limit repair interactions, helping companies
and individual engineers control the cost of AI-assisted development.
Acceptance criteria specify the evidence required for delivery. %
On budget exhaustion, the incomplete outcome records unresolved checks
for engineering review. Internal inference addresses the external
model-service data boundary alongside organizational controls over access
to code, model inputs, and verification records.

\subsection{Open Source and Internal Applications}
\label{sec:reuse}
The open-source edition supports research and teaching through an IDE,
PLC MCP/CLI tools, benchmarks, runtime setup, and
documentation~\cite{semaplcrepo}. Researchers can inspect and extend the
generation--verification loop and evaluate changes using the released
benchmarks.
Source, ladder, variable, and simulation views illustrate the connection
between control logic and execution for teaching. Product-specific
adaptations are maintained separately in the proprietary codebase.

The proprietary adaptation embeds the harness in KUKA's
iiQworks~\cite{kukaiiqworks}, connecting PLC generation and
verification to the product environment. The deployment separates three
integration concerns: product-specific project context, internal model
endpoints, and the shared verification procedure. %
The shared loop supports public experimentation and internal engineering,
with separate proprietary product adapters.

Our earlier demonstration~\cite{tu2026semaplc} presents the IDE's
integrated programming, checking, and visualization capabilities. This
paper studies the project-grounded harness and its completion rule through
two-track comparisons, verification-layer analysis, failure diagnostics,
and resource measurements. MCP and CLI expose verification independently
of the browser, providing a common integration interface for the public
IDE and proprietary product. %
The task corpora build on Agents4PLC and
Spec2Control~\cite{agents4plc,spec2control}, with adaptations and scoring
defined in Section~\ref{sec:setup}.

\subsection{Engineering Impact and Practitioner Feedback}
In our engineering practice, PLC programming and debugging for a typical
custom automation project require about one person-month. We deployed
agent capabilities to support PLC engineers across multiple internal
projects. Project reports indicate an average 45\% reduction in PLC
development time with agent assistance, complementing the benchmark
results with evidence of practical value.

A practitioner described an urgent customer task in his team:

\noindent\textit{``The day before the holiday, an external customer asked
us to develop all the PLC programs and carry out on-site commissioning
after the holiday. I initially expected that we would need to work
overtime. Using the provided agent capabilities, we completed the PLC
code development in one day. We can now use the code on site after the
holiday.''}

This experience illustrates agent-assisted PLC development under a tight
delivery schedule. It provides a practitioner perspective on the reported
efficiency gains.

\subsection{Threats to Validity and Scope}
\label{sec:validity}
\paragraph{Attribution and comparison}
Method comparisons retain different feedback and interaction budgets.
Bare removes skills and tools together, and the ablation uses one backbone
and one layer ordering. These settings measure complete configurations
and conditional increments, leaving the gate's budget-matched effect
unisolated. Repeated-run variability is unmeasured, so small margins are
descriptive. 

\paragraph{Measurement and transfer}
The static behavior score measures satisfaction of selected text assertions,
and formal
verified pass depends on verifier support, including inconclusive timer
results. Runtime scores include execution failures and use exact sampled
values under bounded scenarios. Unscored alarm/fault ports and transients
without retained effects limit coverage. Code and generated assertions
can share interpretation errors, motivating independent scoring oracles.
The 65 plant-section tasks use one evaluation setup. Different scan
timing, I/O latency, and vendor toolchains require renewed validation.
Product integration supplies the application context, while benchmark
scores describe the evaluated configuration. Operational acceptance
proceeds through plant-specific tests and engineer-led commissioning.

\section{Conclusion}
\label{sec:conclusion}
We present SemaPLC, a project-grounded, verification-gated agent harness for
AI-assisted PLC code generation. Project context and reusable skills guide
implementation, while specification review, compilation, and live-runtime
checks drive repair and govern delivery. Two evaluation
tracks connect independent-POU verification with project integration and
runtime behavior. Across seven backbones, SemaPLC achieves the highest
verified-pass and dynamic scores among the evaluated methods. Layer and
failure analyses distinguish executable delivery from output agreement,
and cost measurements characterize the resources required. %
Industrial product integration supports engineering within the company,
while the open-source IDE supports research and teaching. Internal
deployment reports an average 45\% reduction in PLC development time.
The central engineering
contribution is a development process that delivers project-compatible
control logic together with current verification evidence.

\section*{Data Availability}
\label{sec:availability}
SemaPLC are publicly
available in the following repository:
\par
{\raggedright\noindent
\url{https://github.com/midea-ai/SemaPLC}\par}
\noindent
The release includes the open-source IDE, PLC tools, runtime
setup instructions, and both evaluation benchmarks.

\newcommand{\BIBdecl}{%
  \setlength{\parfillskip}{0pt plus 0.75\linewidth}}
\bibliography{refs}
\bibliographystyle{IEEEtran}

\end{document}